\documentclass[amsmath,amssymb,aps,showkeys,showpacs,pra,preprint]{revtex4-1}

\usepackage[english]{babel}
\usepackage{graphicx}
\usepackage{graphics}
\usepackage{amsmath}
\usepackage{dcolumn}
\usepackage{amssymb}
\usepackage{bm}
\usepackage[latin1]{inputenc}
\usepackage{graphicx,color}
\usepackage[caption=false]{subfig} 

\begin{document}
\title{Landau Quantization for $ \Lambda $-Type Neutral Atoms in an Homogeneous Spin-Dependent Gauge Potential}
\author{B. Farias}
\email{bruno.farias@ufcg.edu.br}
\affiliation{Centro de Ci\^encias e Tecnologia Agroalimentar, Universidade Federal de Campina Grande, 58840-000, Pombal, PB, Brazil.}

\author{C. Furtado}
\email{furtado@fisica.ufpb.br} 
\affiliation{Departamento de F\'isica, Universidade Federal da Para\'iba, Caixa Postal 5008, 58051-970, Jo\~ao Pessoa, PB, Brazil.} 

\begin{abstract}

We investigate the quantum dynamics of neutral atoms subject to a uniform spin-dependent gauge field. In particular, we analyze a simple experimental scheme to generate the Landau quantization in a two dimensional atomic gas with internal three-level $\Lambda$-type configuration. We show how energy eigenfunctions and eigenvalues are obtained and discuss the experimental conditions for which a variety of physical quantities of the atomic gas can exhibit quantum oscillations.

\end{abstract}


\maketitle
\section{Introduction}\label{sec1}
Neutral atomic systems, in contrast with their electronic counterparts, offer unprecedented possibilities of controlling over physical parameters \cite{1,2,3,4,4a}. In this scenario, ultracold atoms can be used as quantum simulators for a wide variety of phenomena \cite{5,6,7,8,9}. Two-dimensional (2D) systems of atoms in synthetic gauge fields are of particular interest \cite{10,11,12,13}. The essential requirement for the emergence of the synthetic magnetism is that the wave function of a neutral particle acquires a geometrical phase when it follows a closed path. This demand is satisfied when the particle moves in an appropriate designed external field configuration. There exists a variety of proposed ways to realize artificial magnetic fields for atoms in a trap, namely: using the coupling between the dipole moment of the atom and a properly designed electromagnetic field arrangement \cite{14,15,16,17}, by the rotation of the trap \cite{18}, by atoms in optical lattices using laser-assisted state sensitive tunnelling \cite{19}, and employing schemes based on the adiabatic motion of atoms, with $\Lambda$ \cite{20} or tripod \cite{21} configurations, in spatially varying laser fields. Particularly interesting the optical methods allow the possibility to create artificial magnetic fields with Abelian and/or non-Abelian structure.

The motion of neutral particles in a plane and in the presence of a perpendicular synthetic magnetic field is highly special. In the case of a uniform Abelian U(1) magnetic field, the cyclotron motion of the particles leads to the Landau level band structure \cite{22,23,24,25,25a} similar to the electronic systems, with each level providing a macroscopic number of one-particle states that are strictly degenerate in energy. However, the artificial magnetism also allows the generation of an homogeneous magnetic field U(1) x U(1) that acts in opposite directions on electrically neutral atoms with opposite spin polarizations. As a consequence, it is possible to study many effects related to the Landau quantization exploring the spin-dependence of the effective magnetic field. In Ref. \cite{13} we have proposed a laser configuration to create degenerate Landau levels for tripod-type cold atoms in the presence of a spin-dependent optically induced magnetic field. For $\Lambda$-type atoms the spin-dependent gauge fields, which lead to spin-orbit (SO) coupling, was proposed in \cite{26}. Other potential application of synthetic magnetism is the realization of the Landau quantization for ultra-cold atoms in presence of non-Abelian magnetic fields. Theoretical works show that non-Abelian effects are responsible to break the energy degeneracy and to strongly modify the Landau level structure \cite{27,28}. 

In the present paper, we propose an experimental scheme to study the Landau quantization for a two-dimensional (2D) atomic gas with a simple internal $\Lambda$-type setup submitted to a uniform spin-dependent magnetic field. In comparison to the tripod configuration \cite{21}, we need a large detuning in this $ \Lambda $ configuration. However, the large detuning regime makes the pseudospins of the $\Lambda$ system more stable under atomic spontaneous emission since they have negligible contribution from the initial excited state, while the pseudospins in the tripod system are not the ground states. In addition, the laser beam arrangement is simpler in the $\Lambda$-level setup. The energy eigenfunctions and eigenvalues of this Landau problem are obtained. We discuss about the experimental conditions to reach the strong magnetic field regime. In this regime we would expect the appearance of an oscillatory behaviour of some physical observables (conductivity, magnetization, specific heat etc.) as a function of the magnetic field strength.
\section{The Three-level $\Lambda$-type configuration}\label{sec2}
\begin{figure}[!h]
	\begin{center}
		\subfloat[]{\label{fig_1a}\includegraphics[scale= 0.8]{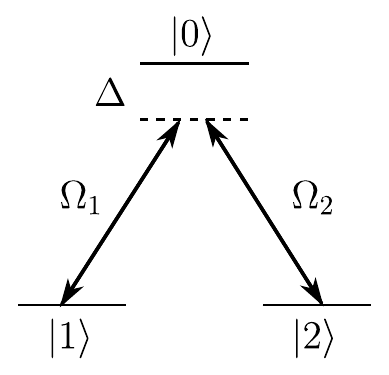}}
			\hspace{0.5cm}
		\subfloat[]{\label{fig_1b}\includegraphics[scale= 0.4]{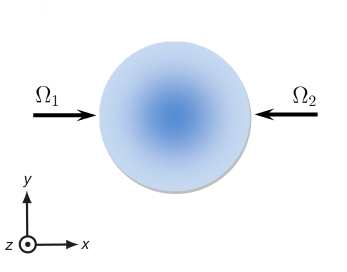}}
	\end{center}
\caption{(a) The $ \Lambda $ configuration, when the Rabi frequencies are parametrized as $ \Omega_1 $ and $ \Omega_2 $ with $ \Delta $ being the detuning. (b) Two counterpropagating beams interacting with the two-dimensional atomic cloud formed by $ \Lambda $-type atoms.}
\end{figure}

In this section, we consider a cloud of an 2D (x-y plane) atomic gas with internal  $\Lambda$ level structure coupled to laser fields [see Fig. (\ref{fig_1a})]. The two ground states $ |1\rangle $ and $ |2\rangle $ are coupled to the excited state $ |0\rangle $ through spatially varying laser fields, with Rabi frequencies $ \Omega_1 $ and $ \Omega_2 $, respectively. The total wave function $ |\Psi({\bf r}) \rangle=\sum_{j=1}^{3} \psi_j ({\bf r}) |j\rangle $ of the atom, where $ {\bf r} $ denotes the atomic position, is governed by the total Hamiltonian $ H=\frac{{p}^2}{2m}+V({\bf r})+ H_{I} $, where $ {\it m} $ is the atomic mass.  

The external trapping potential $ V({\bf r})=\sum_{j} V_j ({\bf r}) |j\rangle \langle j| $ is diagonal in the basis of the internal states $ |j\rangle $, and the laser-atom interaction Hamiltonian $ H_{I} $ is given by
\begin{equation}
H_{I}=\hbar\left[
\Delta|0\rangle\langle 0|
+\Omega_1 |0\rangle\langle 1|
+\Omega_2 |0\rangle\langle 2|
+H.c.
\right]
\end{equation}
with $ \Delta $ as the detuning. As in the reference \cite{29} we parametrize two Rabi frequencies through $ \Omega_1=\frac{\Delta}{2}\tan2\theta\cos\varphi e^{iS_1} $ and $ \Omega_2=\frac{\Delta}{2}\tan2\theta\sin\varphi e^{iS_2} $ with $ -\pi/4 < \theta < \pi/4 $.

Diagonalizing the interacting Hamiltonian $ H_{I} $ yields the eigenstates:
$ |\chi_1 \rangle = \sin \varphi e^{-iS_1}|1\rangle -\cos\varphi e^{-iS_2}|2\rangle $, 
$ |\chi_2 \rangle = \cos\theta ( \cos \varphi e^{-iS_1}|1\rangle +\sin\varphi e^{-iS_2}|2\rangle )-\sin\theta |0\rangle $ and
$ |\chi_3 \rangle = \sin\theta ( \cos \varphi e^{-iS_1}|1\rangle +\sin\varphi e^{-iS_2}|2\rangle )+\cos\theta |0\rangle $. The corresponding eigenvalues are $ E_1=0 $, $ E_2=-\hbar\Delta\frac{\sin^2\theta}{cos(2\theta)} $ and $ E_3=\hbar\Delta\frac{\cos^2\theta}{cos(2\theta)} $.

In the new basis $ |\chi\rangle $, the full quantum state is represented as $ |\Psi({\bf r}) \rangle=\sum_{j=1}^{3} \tilde{\psi}_j ({\bf r}) |\chi_j ({\bf r})\rangle $, where the wave functions $ \tilde{\psi}=(\tilde{\psi}_1, \tilde{\psi}_2, \tilde{\psi}_3)^{Tr} $ obey the Schr\"odinger equation $ i\hbar \partial_t \tilde{\psi} = H_{\mathrm{eff}} \tilde{\psi} $, with the effective Hamiltonian $ H_{\mathrm{eff}}=\frac{({\mathbf{p}}-\mathbf{A})^2}{2m}+\tilde{V} $. Here $ \bf{A} $ and $\tilde{V} $ are $ 3 \times 3 $ matrices with elements given by
\begin{equation}\label{Eq1}
\mathbf{A}_{j,l} = i\hbar \langle\chi_j |{\bf \nabla} \chi_l\rangle,
\end{equation} 
\begin{equation}\label{Eq2}
\tilde{V}_{j,l} = E_j \delta_{j,l} + \langle\chi_j|V({\bf r})|\chi_l\rangle.
\end{equation}

For our purpose, we set that $\Delta > 0 $ and $ -\pi/4 < \theta < \pi/4 $, so there are $ E_3 > E_1 \geq E_2 $ and $ E_3 -E_1 \geq \hbar \Delta $. Since $ \mathbf{A}_{j,l} $ usually has a magnitude of momentum $ P_L $ of the applied lasers fields, the off-diagonal elements of $ \mathbf{A} $ and $ \tilde{V} $ can be neglected when $ \frac{P^2_L}{2m} << \hbar \Delta $ and $ \langle\chi_j|V({\bf r})|\chi_l\rangle $ and the atoms move very slowly (i.e., $ \frac{(\mathbf{p})^2}{2m} << \hbar \Delta $). Applying these adiabatic conditions, the state $|\chi_3 \rangle$, whose eigenenergy $ E_3 $ is much larger than the other eigenenergies $ E_1 $ and $ E_2 $, is decoupled from the other lower internal eigenstates. Thus, we have a near-degenerate subspace spanned by the two lower internal eigenstates {$ \chi_1 $, $ \chi_2 $}. This provides an effective spin-$1/2$ system where the pseudospin up and down states are represented by $ |\uparrow \rangle \equiv |\chi_1 \rangle $ and $ |\downarrow \rangle \equiv |\chi_2\rangle $, respectively. In this way, the wave function $ \Psi $ evolutes under the effective Hamiltonian
\begin{equation}
H_{\mathrm{eff}}=\frac{(\mathbf{p}-\mathbf{A})^2}{2m}+\tilde{V}+\Phi,
\end{equation}
where $ \mathbf{A} $ and $ \tilde{V} $ are truncated $ 2 \times 2 $ matrices with elements defined in Eqs. (\ref{Eq1}) and (\ref{Eq2}) and $ j,l = 1,2 $. In addition, the scalar potential $ \Phi $, which is again a $ 2 \times 2 $ matrix, has elements $ \Phi_{j,l} = \frac{1}{2m} \mathbf{A}_{j,3} \cdot \mathbf{A}_{3,l} $. Using the expression of the eigenstates $ |\chi_1 \rangle $ and $ |\chi_2\rangle $ we can written
\begin{eqnarray}
\mathbf{A}_{11}&=&\hbar\left(\cos^2\varphi\nabla S_{2} + \sin^2\varphi \nabla S_{1}\right),
\nonumber\\
\mathbf{A}_{12}&=&\hbar\cos\theta\left[\frac{1}{2} \sin\left(2\phi\right)\nabla(S_1-S_2) 
- i \nabla \varphi\right],
\nonumber\\
\mathbf{A}_{22}&=&\hbar\cos^2\theta\left(\cos^ 2\varphi\nabla S_{1} + \sin^ 2\varphi\nabla S_{2}\right),
\nonumber
\end{eqnarray}

\begin{eqnarray}
\Phi_{11}
&=&\frac{\hbar^2}{2m}\sin^ 2\theta\left[\frac{1}{4}\sin^2(2\varphi)(\nabla S_1 -\nabla S_2)^2
+(\nabla \varphi )^2\right],
\nonumber\\
\Phi_{12}
&=&
\frac{\hbar^2}{2m}\sin\theta
\left[\frac{1}{2}\sin(2\varphi)\nabla( S_1 - S_2)
-i\nabla\varphi\right]
\nonumber\\
&&\cdot
\left[\frac{1}{2}\sin(2\theta)
\left(\cos^ 2\varphi\nabla S_1
+\sin^2\varphi \nabla S_2\right) -i\nabla\theta \right],
\nonumber\\
\Phi_{22}
&=&
\frac{\hbar^2}{2m}\left[
\frac{1}{4}\sin^2(2\theta)
\left(\cos^ 2\varphi\nabla S_1
+\sin^2\varphi \nabla S_2\right)^2
+(\nabla\theta)^2 \right].
\nonumber
\end{eqnarray}
and
\begin{eqnarray}
\tilde{V}_{11}
&=&
\left(V_1\sin^2\varphi  + V_2\cos^2\varphi \right),
\nonumber\\
\tilde{V}_{12}
&=&
\left(V_1 - V_2\right)\cos\theta\frac{\sin(2\varphi)}{2},
\nonumber\\
\tilde{V}_{22}
&=&
E_2+\cos^2\theta\left(V_1\cos^2\varphi  +V_2\sin^2\varphi  \right) + V_3\sin^2\theta. 
\nonumber
\end{eqnarray}

We now consider a specific configuration of two contrapropagating plane waves laser beams in the $\mathbf{x}$ direction for the formation of the Landau quantization in the atomic gas. The spatial profiles of the corresponding Rabi frequencies are assumed to be of planes waves form [see Fig. (\ref{fig_1b})]
\begin{equation}
\Omega_1=\tilde{\Omega}(\theta) e^{i\kappa x}
\quad\mbox{and}\quad
\Omega_2=\tilde{\Omega}(\theta) e^{-i\kappa x},
\end{equation}
with $\tilde{\Omega}(\theta)=\frac{\Delta}{2}\tan(2\theta) $. Note that the phases are $ S_1=\kappa x $, $ S_2=-\kappa x $ and the angles are $ \varphi =\pi/4 $ and $ \cos\theta = \alpha y $, where $\alpha$ is a parameter.  

Under this laser arrangement, the vector takes the form of the uniform $ U (1) \times U(1) $ gauge potential  

\begin{equation}
\mathbf{A}=
\left(\begin{array}{cc}
0  & B y \mathbf{x} \\
B y \mathbf{x} & 0 \\
\end{array}\right)= B y \sigma_x \mathbf{x},
\end{equation}
where $ B =\hbar\kappa\alpha $. Since the vector potential has $ U (1) \times U(1) $ structure, so the dynamics of atoms in any pseudospin states evolves according to a separate Hamiltonian \cite{30,31}.

This effective vector potential corresponds to a constant magnetic field in the direction of $z$ axis, since we have the relation
\begin{equation}\label{magnetic_field}
\mathbf{B} =  \nabla \times \mathbf{A} = - B \sigma_x \mathbf{z}.
\end{equation}

The overall trapping potential can expressed in form

\begin{widetext}
\begin{equation}
\Phi + \tilde{V} =
\left(\begin{array}{cc}
V_1 + \frac{\hbar^2\kappa^2}{2m}\sin^ 2\theta  & 0 \\
0 & V_1 \cos^2\theta + V_3 \sin^2\theta+\frac{\hbar^2\alpha^2}{2m}\frac{1}{\sin^2\theta} -\hbar\Delta\frac{\sin^2\theta}{\cos(2\theta)} \\
\end{array}\right),
\nonumber
\end{equation}
\end{widetext}
where the trapping potential is assumed to be the same for the first two atomic states, $V_1 = V_2$.

In order to avoid the spontaneous decay for realistic ultracold atoms, we consider the large detuning case, i.e.,  $ |\Delta| \gg |\Omega_1| $, $ |\Delta| \gg |\Omega_2| $. This leads to the condition $ \sin^2\theta << 1 $ and as a consequence $ \tilde{\Omega}(\theta) \approx \frac{\Delta}{2}\sin(2\theta) $. In this regime  
\begin{equation}
-\hbar\Delta\frac{\sin^2\theta}{1-2\sin^2\theta}
\approx -\hbar\Delta\sin^2\theta
\nonumber
\end{equation}
and the overall trapping potential can be written as
\begin{widetext}
\begin{equation}
\Phi + \tilde{V} =
\left(\begin{array}{cc}
V_1 + \frac{\hbar^2\kappa^2}{2m}\sin^ 2\theta  & 0 \\
0 & V_1 \cos^2\theta + V_3 \sin^2\theta+\frac{\hbar^2\alpha^2}{2m}\frac{1}{\sin^2\theta} -\hbar\Delta \sin^2\theta \\
\end{array}\right).
\nonumber
\end{equation}
\end{widetext}

If we also assume that $ V_{3} - V_{1}= \hbar\Delta -\frac{\hbar^2\alpha^2}{2m}\frac{1}{\sin^4\theta} $ and recall that $ \frac{\hbar^2\kappa^2}{2m}<<\hbar\Delta$, we finally have
\begin{equation}
\Phi + \tilde{V} 
=V_1\mathbb{I},
\nonumber
\end{equation}
where $\mathbb{I}$ is the unit matrix.
\section{Landau Levels for  $ \Lambda $-Type Neutral Atoms}\label{sec3}
In this context, the effective Hamiltonian of the system takes the form
\begin{eqnarray}
H_{\mathrm{eff}}
&=&
\frac{1}{2m}\left(p_{x}-B y \sigma_x\right)^2
+\frac{1}{2m} p_y^2 + V_1\mathbb{I}.
\nonumber
\end{eqnarray}

It is convenient to apply the local unitary transformation $ U=e^{\frac{i\pi}{4}\sigma_y} $ on near-degenerate basis. With this transformation the vector and scalar potentials are $ \mathbf{A'}= B y\sigma_z \mathbf{x} $ and $ \Phi'=\Phi $, and new two-component wave function is related to the original one according to $ \tilde{\psi'} =e^{i\frac{\pi}{4}\sigma_y} \tilde{\psi} $. In addition, using the spin language with $ |\uparrow \rangle \equiv |\chi_1 \rangle $ (pseudospin up), $ |\downarrow \rangle \equiv |\chi_2\rangle $ (pseudospindown) and $ \psi'=(\psi'_{\uparrow}, \psi'_{\downarrow})^{Tr} $, the effective diagonal Hamiltonian takes the form of
\begin{equation}
H'_{\mathrm{eff}} =
\left(\begin{array}{cc}
H'_{\uparrow} & 0 \\
0 & H'_{\downarrow} \\
\end{array}\right)
\nonumber
\end{equation}
where $ H'_{\gamma} = \frac{1}{2m}\left(p_{x}-\mathbf{A'}_{\gamma}\right)^2+\frac{1}{2m} p_{y}^2+V_1\mathbb{I} $, ($ \gamma = \uparrow, \downarrow $) and $ \mathbf{A'}_{\uparrow} = - \mathbf{A'}_{\downarrow} = B y \mathbf{x} $.

The transformed Schr\"odinger equation of the atomic motion in the pseudospin-1/2 basis  $ \{|\uparrow \rangle ,|\downarrow \rangle \} $ yields a system of two decoupled equations
\begin{eqnarray}
\label{Eq3}
\Bigg[
\frac{1}{2m}p_{y}^2
+\left(
\frac{B^2}{2m}
\right)y^2
&&
-\eta_{\gamma}\frac{B\hbar k_x}{m}y
\nonumber\\&&
+\frac{\hbar^2 k_x^2}{2m}+V_1
\Bigg]
\psi'_{\gamma} 
= 
E \psi'_{\gamma},
\end{eqnarray}
where $ \eta_{\uparrow} = - \eta_{\downarrow} = 1 $.

Since the transverse momentum $p_x=\hbar k_x $ is the quantum integral of motion so that the solution $\psi'_{\gamma} $ can be factorized to separate the variables
\begin{equation}\label{Eq4}
\psi'_{\gamma}(x,y)
=
e^{ik_x x}\psi'_{\gamma}(y);
\quad
\psi'_{\gamma}(y) 
=
\left(\begin{array}{c}
\psi'_{\uparrow}(y) \\ \psi'_{\downarrow}(y) 
\end{array}\right).
\end{equation}

Substituting the plane wave ansatz (\ref{Eq4}) into Eq.(\ref{Eq3}) and completing the square for the variable $y$ one arrives in two spin-dependent harmonic oscillators with their well centred at $ y^{{\gamma}}_{0} = \frac{\eta_{\gamma}\hbar k_x}{B} $.
\begin{eqnarray}\label{Eq5}
\Bigg[\frac{1}{2m}p_{y}^2
+\frac{B^2}{2m}
\left(
y
-y^{{\gamma}}_{0}
\right)^2
\Bigg]\psi'_{\gamma}(y)
=
\Bigg[E -V_1\Bigg]\psi'_{\gamma}(y).
\end{eqnarray} 

Thus, the effective magnetic field leads to a Landau level structure for each spin state. In addition, the spin-hall effect would be realized in a situation where the atomic trap is turned off and the atoms fall due to the gravity.

By applying the change of variables $ y_{\gamma} = y-y^{{\gamma}}_{0} $ the equations (\ref{Eq5}) takes the form of the ordinary harmonic oscillator equations
\begin{eqnarray}\label{Eq6}
-\frac{\hbar^2}{2m}\frac{d^2\psi'_{\gamma}(y)}{dy_{\gamma}^2}
+\frac{m\omega^2}{2}y_{\gamma}^2\psi'_{\gamma}(y)
&=&\epsilon\psi'_{\gamma}(y),
\end{eqnarray}
where
$ \omega = \frac{B}{m} $ and
\begin{equation}\label{energy1}
\epsilon = E-V_1.
\end{equation}

Introducing the dimensionless variable $ \xi_{\gamma} = \sqrt{\frac{m\omega}{\hbar}}y_{\gamma} $ we rewrite Eq. (\ref{Eq6}) in a dimensionless form
\begin{eqnarray}\label{equation6}
\frac{d^2\psi_{\gamma}}{d\xi_{\gamma}^2}=\left( \xi_{\gamma}^2-\frac{2\epsilon}{\hbar\omega} \right) \psi_{\gamma}
\end{eqnarray} 
  
At this point, it is important to remember that we are considering trapped particles in an atomic cloud. In this way, we cannot take the asymptotic limit of the variable $ y_{\gamma} $. However, we obtain analytical solutions of Eq. (\ref{equation6}) in the limit  $\sqrt{\frac{m\omega}{\hbar}}\gg 1$ \cite{20}. Note that both sides of such inequality have dimensions of inverse length. As a consequence, the magnetic field strength must obey the following condition
\begin{equation}\label{condition3}
\sqrt{\kappa\alpha} \gg 1\ \mathrm{and}\ \ \sqrt{\frac{B}{\hbar}} \gg 1.
\end{equation} 

Solving the differential equation (\ref{equation6}) we obtain a family of solutions in the form of $ \psi_{n}(\xi_{\gamma})=\frac{1}{\sqrt{2^nn!}}\left(\frac{m\omega}{\pi\hbar}\right)^{1/4} e^{-\frac{m\omega} {2\hbar}\xi_{\gamma}^2}H_n\left(\sqrt{\frac{m\omega}{\hbar}}\xi_{\gamma} \right) $ \cite{32}, with $H_n$ the usual Hermite polynomial. In addition, $\epsilon$ can assume the values
\begin{equation}\label{energy2}
\epsilon_n = \hbar\omega\left(n+\frac{1}{2}\right),
\end{equation}
with $ n = 0,1,2,... $ . Thus, comparing Eq.(\ref{energy1}) and Eq.(\ref{energy2}), one arrives that the energy levels of the two-dimensional atomic gas are quantized into Landau levels
\begin{eqnarray}\label{energy3}
E_{n,{\uparrow}}=E_{n,{\downarrow}}&=&\hbar\omega\left(n+\frac{1}{2}\right)
+V_1,
\end{eqnarray}
where $ \omega = \frac{B}{m} $ is the cyclotron frequency. If the trapping potential $V_{1}$ is constant this additional term only shifts the energy spectrum. The spin-degenerate energy levels are equally spaced by $ \hbar\omega $. The effective magnetic length in this system is $ l = \sqrt{\frac{\hbar}{B}} = \sqrt{\frac{1}{\kappa\alpha}} $ with $ A $ being the area of the trap. $N_{\phi} =  \frac{A}{2\pi l^{2}}$ is the degeneracy of each Landau level. $ \nu = \frac{N}{N_{\phi}} $ is the filling factor, where $N$ is the number of atoms into the atomic cloud.

The degeneracy $N_{\phi}$ is linear in strength of the magnetic field. Decreasing the magnetic field provokes $N_{\phi}$ to decrease, and fewer atoms can be accommodated on each level. As a result, the atomic population of the highest energy level will range from to entirely empty to completely full. When the filling number is of the order of unity, we expect the atomic gas to exhibit an oscillatory dependence on physical observables as a function of field strength at low temperature. Generally known as "quantum magnetic oscillations", these effects could be observed for example, for the atomic analog of the magnetization (de Haas-van Alphen oscillations) \cite{33,33a}, the resistivity (Shubnikov-de Haas oscillations) \cite{34}, the Hall resistance \cite{35}, or the specific heat \cite{36}.

For a typical 2D ultracold atomic gas with an effective area $ A \sim 10000\ \mathrm{{\mu m}^2} $ containing $ N \approx 10^{4} $ atoms of $ ^{87}\mathrm{Rb}  $ \cite{37}, with the wave number $ \kappa_{\mathrm{Rb}} \sim 10^7\ \mathrm{m}^{-1} $ the degeneracy of each Landau level is then estimated by $N_{\phi} = 1.51 \times 10^{25} B $. In addition, from Eq. (\ref{condition3}) $ \alpha \gg 10^{-7}\ \mathrm{m}^{-1} $ and $ B \gg 10^{-14}\hbar\kappa_{\mathrm{Rb}}^2 $. In this context, the lowest Landau level regime ($ \nu = 1 $) is reached when $ B = 6.28 \times 10^{-2}\hbar\kappa_{\mathrm{Rb}}^2 $. 
\section{Concluding Remarks}\label{sec4}
To summarize, we have investigated the analog of Landau levels by considering a uniform spin-dependent magnetic field induced in a 2D ultracold atomic gas with $\Lambda$-type configuration of internal states. The strength of the effective magnetic field depends on the relative intensity of the laser beams interacting with the neutral atoms and is limited by the finite size of the atomic cloud. The energy eigenfunctions and eigenvalues of the system are obtained. We have estimated the values of the physical parameters to achieve the regime in which the particles populate just the lowest Landau level.

\acknowledgments{This work was support by the Brazilian agencies CNPq , CAPES and FAPESQ. Helpful discussions with J. Lemos de Melo are gratefully acknowledged}.


\begin{thebibliography}{99}

\bibitem{1} B. DeMarco and D. S. Jin, Science {\bf 285}, 1703 (1999).

\bibitem{2} M.-O. Mewes, G. Ferrari, F. Schreck, A. Sinatra, and C.Salomon, Phys. Rev. A {\bf 61}, 011403(R) (2000).

\bibitem{3} L. Pitaevskii and S. Stringari, {\it Bose-Einstein Condensation} (Clarendon Press, Oxford, 2003).

\bibitem{4} B. Farias, T. Passerat de Silans, M. Chevrollier and M. Ori\'a, Phys. Rev. Lett. {\bf 94}, 173902 (2005)
 \bibitem{4a}M. Ori\'a, B. Farias, T. Sorrentino, and M. Chevrollier, J. Opt.Soc. Am. B {\bf 24}, 1867 (2007).

\bibitem{5} J. R. Anglin and W. Ketterle, Nature (London) 416, {\bf 211} (2002).

\bibitem{6} U. Leonhardt, T. Kiss, and P. Ohberg, Phys. Rev. A {\bf 67}, 033602 (2003).

\bibitem{7} D. Jaksch and P. Zoller, New J. Phys. {\bf 5}, 56 (2003).

\bibitem{8} Y. J. Lin {\it et al.}, Nature (London) {\bf 462}, 628 (2009). 

\bibitem{9} G. Juzeli$ \bar{\mathrm{u}} $nas, J. Ruseckas, M. Lindberg, L. Santos and P. \"Ohberg, Phys. Rev. A {\bf 77}, 011802(R) (2008).

\bibitem{10} Indubala I. Satija, Daniel C. Dakin, J. Y. Vaishnav, and Charles W. Clark, Phys. Rev. A {\bf 77}, 043410 (2008).

\bibitem{11} X. -J. Liu, M. F. Borunda, X. Liu, and J. Sinova, Phys. Rev. Lett. {\bf 102}, 046402 (2009).

\bibitem{12} Dan-Wei Zhang, Zheng-Yuan Xue, Hui Yan, Z. D. Wang,2, and Shi-Liang Zhu, Phys. Rev. A {\bf 85}, 013628 (2012).

\bibitem{13} B. Farias, J. Lemos de Melo and C. Furtado, Eur. Phys. J. D {\bf 68}, 77 (2014).

\bibitem{14} Y. Aharonov, and A. Casher, Phys. Rev. Lett. {\bf 53}, 319 (1984).

\bibitem{15} X.-G. He, B.H.J. McKellar, Phys. Rev. A {\bf 47}, 3424 (1993).

\bibitem{16} M. Wilkens, Phys. Rev. Lett. {\bf 72}, 5 (1994).

\bibitem{17} H. Wei, R. Han, X. Wei, Phys. Rev. Lett. {\bf 75}, 2071 (1995).

\bibitem{18} A. L. Fetter, Rev. Mod. Phys. {\bf 81}, 647 (2009).

\bibitem{19} E. J. Mueller, Phys. Rev. A {\bf 70}, 041603(R) (2004). 

\bibitem{20} G. Juzeli$ \bar{\mathrm{u}} $nas and P. \"Ohberg, Phys. Rev. Lett. {\bf 93}, 033602 (2004). 

\bibitem{21} J. Ruseckas, G. Juzeli$ \bar{\mathrm{u}} $nas, P. \"Ohberg, and M. Fleischhauer, Phys. Rev. Lett. {\bf 95}, 010404 (2005).

\bibitem{22} P. \"Ohberg, G. Juzeli$ \bar{\mathrm{u}} $nas, J. Ruseckas, and M. Fleischhauer, Phys. Rev. A {\bf 72}, 053632 (2005).

\bibitem{23} A. L. Fetter, Rev. Mod. Phys. {\bf 81}, 647 (2009).

\bibitem{24} M. Ericsson, E. Sj\"oqvistt, Phys. Rev. A {\bf 65}, 013607 (2001).

\bibitem{25} L.R. Ribeiro, C. Furtado, and J.R. Nascimento, Phys. Lett. A {\bf 348}, 135 (2006).
\bibitem{25a} L.R. Ribeiro, C. Furtado, J.R. Nascimento, Phys. Lett. A {\bf 358}, 336 (2006).

\bibitem{26} X. -J. Liu, M. F. Borunda, X. Liu, and J. Sinova, Phys. Rev. Lett. {\bf 102}, 046402 (2009).

\bibitem{27} A. Jacob, P. \"Ohberg, G. Juzeli$ \bar{\mathrm{u}} $nas, and L. Santos, New J. Phys. {\bf 10}, 045022 (2008).

\bibitem{28} B. Estienne, S. Haaker, and K. Schoutens, New J. Phys. {\bf 13}, 045012 (2011).

\bibitem{29} M. -Y. Ye and X. -M. Lin, e-print arXiv:1207.5369v1.

\bibitem{30} L. S. Brown and W. I. Weisberger, Nucl. Phys. B {\bf 157}, 285-326 (1979).

\bibitem{31} B. Estienne, S. Haaker, and K. Schoutens, New J. Phys. {\bf 13}, 045012 (2011).

\bibitem{32} L. D. Landau, E. M. Lifshitz, {\it Quantum Mechanics: Non-relativistic Theory}, 3rd ed., (Pergamon Press, Oxford, 1977). 

\bibitem{33} B. Farias and C. Furtado, Physica B {\bf 481}, 19 (2016).
 \bibitem{33a}B. Farias and C. Furtado, Eur. Phys. J. Plus {\bf 131}, 237 (2016).

\bibitem{34} Bikash Padhi and Sankalpa Ghosh, Phys. Rev. Lett. {\bf 111}, 043603 (2013).

\bibitem{35} Wolfgang Ketterle, Nature Physics {\bf 11}, 90 (2015).

\bibitem{36} G. Juzeli$ \bar{\mathrm{u}} $nas and P. \"Ohberg, Phys. Rev. Lett. {\bf 93}, 033602-1 (2004).

\bibitem{37} K. Merloti, R. Dubessy, L. Longchambon, A. Perrin, P.-\'{E}. Pottie, V. Lorent, and H. Perrin, New J. Phys. {\bf 15}, 033007 (2013).




\end{thebibliography}
\end{document}